%% file: main.tex
\documentclass[10pt,a4paper,twocolumn,superscriptaddress,aps,prd,longbibliography,nofootinbib]{revtex4-2}
\usepackage{amsmath}
\usepackage{amsfonts}
\usepackage{amssymb}
\usepackage{bbold}
\usepackage{graphicx}
\usepackage[usenames,dvipsnames]{color}
\usepackage{float}
\usepackage{multirow}
\usepackage{soul}
\usepackage[nohyperlinks, printonlyused, nolist]{acronym}

\usepackage{blindtext}

\begin{document}

\author{Cedric Schmitt} 
\thanks{These authors have contributed equally}
\affiliation{Physikalisches Institut, Universit\"at W\"urzburg, D-97074 W\"urzburg, Germany}
\affiliation{W\"urzburg-Dresden Cluster of Excellence ct.qmat, Universit\"at W\"urzburg, D-97074 W\"urzburg, Germany} 

\author{Jonas Erhardt}
\thanks{These authors have contributed equally}
\affiliation{Physikalisches Institut, Universit\"at W\"urzburg, D-97074 W\"urzburg, Germany}
\affiliation{W\"urzburg-Dresden Cluster of Excellence ct.qmat, Universit\"at W\"urzburg, D-97074 W\"urzburg, Germany}

\author{Philipp Eck}
\affiliation{W\"urzburg-Dresden Cluster of Excellence ct.qmat, Universit\"at W\"urzburg, D-97074 W\"urzburg, Germany}
\affiliation{Institut f\"ur Theoretische Physik und Astrophysik, Universit\"at W\"urzburg, D-97074 W\"urzburg, Germany}

\author{Matthias Schmitt}
\affiliation{Physikalisches Institut, Universit\"at W\"urzburg, D-97074 W\"urzburg, Germany}
\affiliation{Diamond Light Source, Harwell Science and Innovation Campus, Didcot, OX11 0DE, United Kingdom}

\author{Kyungchan Lee}
\affiliation{Physikalisches Institut, Universit\"at W\"urzburg, D-97074 W\"urzburg, Germany}
\affiliation{W\"urzburg-Dresden Cluster of Excellence ct.qmat, Universit\"at W\"urzburg, D-97074 W\"urzburg, Germany}

\author{Tim Wagner}
\affiliation{Physikalisches Institut, Universit\"at W\"urzburg, D-97074 W\"urzburg, Germany}
\affiliation{W\"urzburg-Dresden Cluster of Excellence ct.qmat, Universit\"at W\"urzburg, D-97074 W\"urzburg, Germany}

\author{Philipp Ke\ss ler}
\affiliation{Physikalisches Institut, Universit\"at W\"urzburg, D-97074 W\"urzburg, Germany}
\affiliation{W\"urzburg-Dresden Cluster of Excellence ct.qmat, Universit\"at W\"urzburg, D-97074 W\"urzburg, Germany}

\author{Martin Kamp}
\affiliation{Physikalisches Institut, Universit\"at W\"urzburg, D-97074 W\"urzburg, Germany}
\affiliation{Physikalisches Institut and R\"ontgen Center for Complex Material Systems, D-97074 W\"urzburg, Germany}

\author{Timur Kim}
\affiliation{Diamond Light Source, Harwell Science and Innovation Campus, Didcot, OX11 0DE, United Kingdom}

\author{Cephise Cacho}
\affiliation{Diamond Light Source, Harwell Science and Innovation Campus, Didcot, OX11 0DE, United Kingdom}

\author{Tien-Lin Lee}
\affiliation{Diamond Light Source, Harwell Science and Innovation Campus, Didcot, OX11 0DE, United Kingdom}

\author{Giorgio Sangiovanni}
\affiliation{W\"urzburg-Dresden Cluster of Excellence ct.qmat, Universit\"at W\"urzburg, D-97074 W\"urzburg, Germany}
\affiliation{Institut f\"ur Theoretische Physik und Astrophysik, Universit\"at W\"urzburg, D-97074 W\"urzburg, Germany}

\author{Simon Moser}
\affiliation{Physikalisches Institut, Universit\"at W\"urzburg, D-97074 W\"urzburg, Germany}
\affiliation{W\"urzburg-Dresden Cluster of Excellence ct.qmat, Universit\"at W\"urzburg, D-97074 W\"urzburg, Germany}

\author{Ralph Claessen}
\email{e-mail: claessen@physik.uni-wuerzburg.de}
\affiliation{Physikalisches Institut, Universit\"at W\"urzburg, D-97074 W\"urzburg, Germany}
\affiliation{W\"urzburg-Dresden Cluster of Excellence ct.qmat, Universit\"at W\"urzburg, D-97074 W\"urzburg, Germany}

\date{\today}


\title{Stabilizing an atomically thin quantum spin Hall insulator at ambient conditions: Graphene-intercalation of indenene}

\maketitle

\input{Abstract}
\input{Introduction}

\input{ResultsandDiscussion}

{\noindent

	\textbf{Data Availability} 
		The data that support the plots within this paper and other findings of this study are available from the corresponding author upon reasonable request. 
}


{\noindent
	\textbf{Acknowledgements}
	
	We are grateful for funding support from the Deutsche Forschungsgemeinschaft (DFG, German Research Foundation) under Germany’s Excellence Strategy through the Würzburg-Dresden Cluster of Excellence on Complexity and Topology in Quantum Matter ct.qmat (EXC 2147, Project ID 390858490) as well as through the Collaborative Research Center SFB 1170 ToCoTronics (Project ID 258499086).
	We acknowledge Diamond Light Source for time on beamline I09 and I05 under proposals SI31808-1, 
	SI25151-4 
    and SI30583-1. 
	We gratefully acknowledge the Gauss Centre for Supercomputing e.V. (https://www.gauss-centre.eu) for funding this project by providing computing time on the GCS Supercomputer SuperMUC-NG at Leibniz Super- computing Centre (https://www.lrz.de).
	
}

\begin{scriptsize}
\end{scriptsize} 


{\noindent
	\textbf{Author contributions}
	C.S. and J.E. have realized the epitaxial
growth and surface characterization and carried out the photoelectron spectroscopy experiments
and their analysis. P.E. has conceived the theoretical ideas and performed the DFT, Wannier and Berryology calculations. On the experimental side, contributions came from
M.S., K.L., T.W., P.K., M.K., T.K., C.C., T.-L.L., S.M. and R.C., while G.S. gave inputs to the theoretical aspects. R.C. and S.M. supervised this joint project and wrote the manuscript together with all other authors.
}

{\noindent
	\textbf{Competing interests}
	The authors declare no competing interests.
}


\begin{scriptsize}
\end{scriptsize}

\end{document}

%% file: Abstract.tex
\bigskip\noindent{\bf 
Atomic monolayers on semiconductor surfaces represent a new class of
functional quantum materials at the ultimate two-dimensional limit, ranging from
superconductors \cite{Zhang2010,Ming2023} to Mott insulators
\cite{Profeta2007,Li2013} and ferroelectrics \cite{Guo2023} to quantum spin Hall
insulators (QSHI) \cite{Reis,Bauernfeind}. A case in point is the recently
discovered QSHI indenene \cite{Bauernfeind,PhilippEck2022}, a triangular
monolayer of indium epitaxially grown on SiC(0001), exhibiting a $\sim120$\;meV
gap and substrate-matched monodomain growth on the technologically relevant
$\mu$m scale \cite{Erhardt}. Its suitability for room-temperature spintronics
is countered, however, by the instability of pristine indenene in air, which
destroys the system along with its topological character, nullifying hopes of
\textit{ex-situ} processing and device fabrication. Here we show how indenene
intercalation into epitaxial graphene offers effective protection from the
oxidizing environment, while it leaves the topological character fully intact.
This opens an unprecedented realm of \textit{ex-situ} experimental
opportunities, bringing this monolayer QSHI within realistic reach of actual
device fabrication and edge channel transport. 
}

%% file: Introduction.tex
\bigskip

With the promise of dissipation-less spin-polarized boundary modes, QSHIs could
initiate a paradigm shift in future spintronics technologies. The conceptual
application perspective is bright and ranges from spin-transistors
\cite{Spintronik1, Spintronik2}, to low-power consumption devices
\cite{LowPower2, LowPower1}, to QSHI based quantum computing
\cite{QuantenComp1}. However, finding suitable materials for practicable device
realization faces major challenges. The band-inverted narrow-gap semiconductors
for which the quantum spin Hall effect had first been demonstrated
\cite{Koenig2007, AlSb} do not lend themselves to room-temperature (RT) applications.
2D Dirac semimetals formed by atomic honeycomb monolayers as motivated by the
seminal work of Kane and Mele \cite{Kane2005} are a promising alternative
\cite{Lodge2021}. But while spin-orbit coupling (SOC) in graphene is too weak to open
an appreciable band gap, QSHI monolayers built from heavier group IV elements
such as silicene, germanene, and stanene \cite{Liu2011, Xu2013} could
experimentally only be synthesized on metal surfaces, preventing their use in actual
transport devices, or failed to display a large non-trivial bandgap when placed
on a semiconducting substrate \cite{Kou2017}.

In contrast, band-inverted large gap 2D Dirac semimetals have been
successfully realized in group III and V monolayers on SiC(0001), specifically bismuthene
\cite{Reis} and the recently discovered indenene \cite{Bauernfeind}, and
were experimentally confirmed as QSHIs. They could potentially solve the device
challenge, yet, are inherently unstable to environmental factors outside their
ultra-high vacuum (UHV) birthplace. This bottleneck has made serious transport
device fabrication strategies elusive, and hitherto characterization is mostly
bound to UHV-based surface science techniques such as angle resolved
photoemission (ARPES) and scanning tunneling microscopy (STM).


\begin{figure*}[ht]
\includegraphics[width=17.94cm]{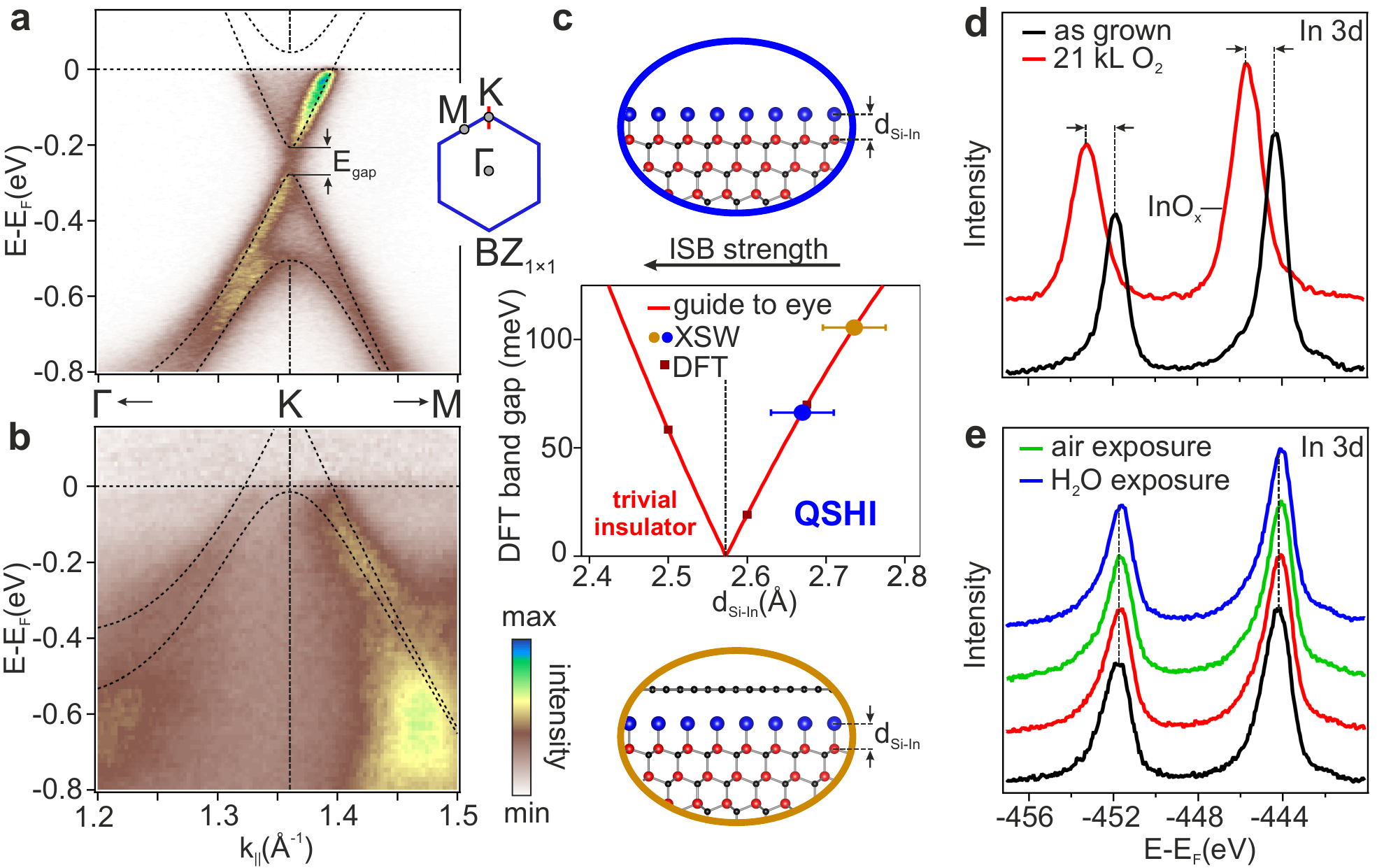}
\caption{\textbf{
Graphene intercalated indenene is topologically non-trivial and resilient to
atmosphere.} \textbf{a,b} ARPES spectra at the $\text{K}$-point of
pristine \textbf{a} and intercalated \textbf{b} indenene (normalized by
integrated EDC) measured with 21.2$\,$eV photons at 20$\,$K and
overlaid DFT (HSE06) calculations (dotted lines) of pristine indenene.
The band splitting in \textbf{a} and \textbf{b} originate from the combined role of SOC and ISB as discussed in the text.
\textbf{c} DFT (HSE06) band gap calculations as well as a guide to the
eye (red line) of pristine indenene as a function of the In-Si bond length
d$_{\text{In-Si}}$, the latter controlling the ISB strength as indicated by a black
arrow. Experimentally (by X-ray standing wave photoemission) determined d$_{\text{In-Si}}$ of
intercalated (yellow data point) and pristine (blue data point) indenene are
placed in this diagram. \textbf{d,e} In $3d$ XPS core-level peaks of \textbf{d} pristine and \textbf{e} intercalated indenene, for the as-grown films (black;
$E_{5/2}=-451.9\,$eV; $E_{3/2}=-444.3\,$eV), after exposure to
21$\,$kL (red) of oxygen (InO$_x$ marks the chemically shifted oxidized In
species), after 10$\,$min exposure to ambient air (green, only \textbf{e}),
and after immersion in liquid water and subsequent mild in-vacuo
degas (blue) as specified in the Methods section.
}
\label{fig:DiracXSWXPS}
\end{figure*}

Here, we design an \textit{extravehicular space suit} to make these quantum
materials operational in air, by placing quasi-freestanding graphene as
protective sheet atop the QSHI monolayer via intercalation. Graphene's
resilience to ambient conditions provides an efficient protection against
oxidation, while it leaves the intercalated material unaffected as was shown for
a variety of few-layer quantum materials \cite{Briggs2020,Wu2021}. With respect
to topological physics, 
intercalation was suggested as a means to tailor the spin-orbit gap of graphene \cite{Li2013B}. In contrast, here we reverse the
roles of graphene and the intercalant, by using the former to stabilize the
latter as a QSHI. For this purpose we employ indenene, the triangular monolayer
phase of indium that can be grown routinely in high quality monodomains on large
areas of SiC \cite{Erhardt}. It is particularly suited for wafer-sized
intercalation, which not only protects both its structure \textit{and} its
topological character but also ensures its chemical integrity upon exposure to
environmental conditions, as we demonstrate below.

In its pristine form, the topological electronic structure of indenene is the result of
a synergetic interplay of the indium monolayer and its underlying SiC
substrate. The latter breaks the surface mirror plane and gaps out the metallic
indium $sp$ states, leaving a set of Dirac bands of in-plane $p$-orbital
character located at the K-point of the Brillouin zone (see
Fig.~\ref{fig:DiracXSWXPS}a). The degeneracy of the Dirac point is lifted by two
counteracting mechanisms \cite{Bauernfeind,PhilippEck2022}. While in-plane
inversion symmetry breaking (ISB), induced by the topmost carbon atoms of the
substrate (see Fig.~\ref{fig:DiracXSWXPS}c), promotes a trivial band gap, atomic SOC drives band inversion and
hence a large QSHI gap. As the bonding
distance d$_{\text{In-Si}}$ between indium and the topmost Si-layer controls the
ISB strength felt by the In monolayer and consequently determines its topology
and gap size, it has been used as \textit{one} among several experimental indicators to
measure indenene's topology. We thus find the pristine form to lie deep within
the QSHI regime (Fig.~\ref{fig:DiracXSWXPS}c, blue data point)
\cite{Bauernfeind}.


\begin{figure*}[ht]
\includegraphics[width=17.94cm, keepaspectratio]{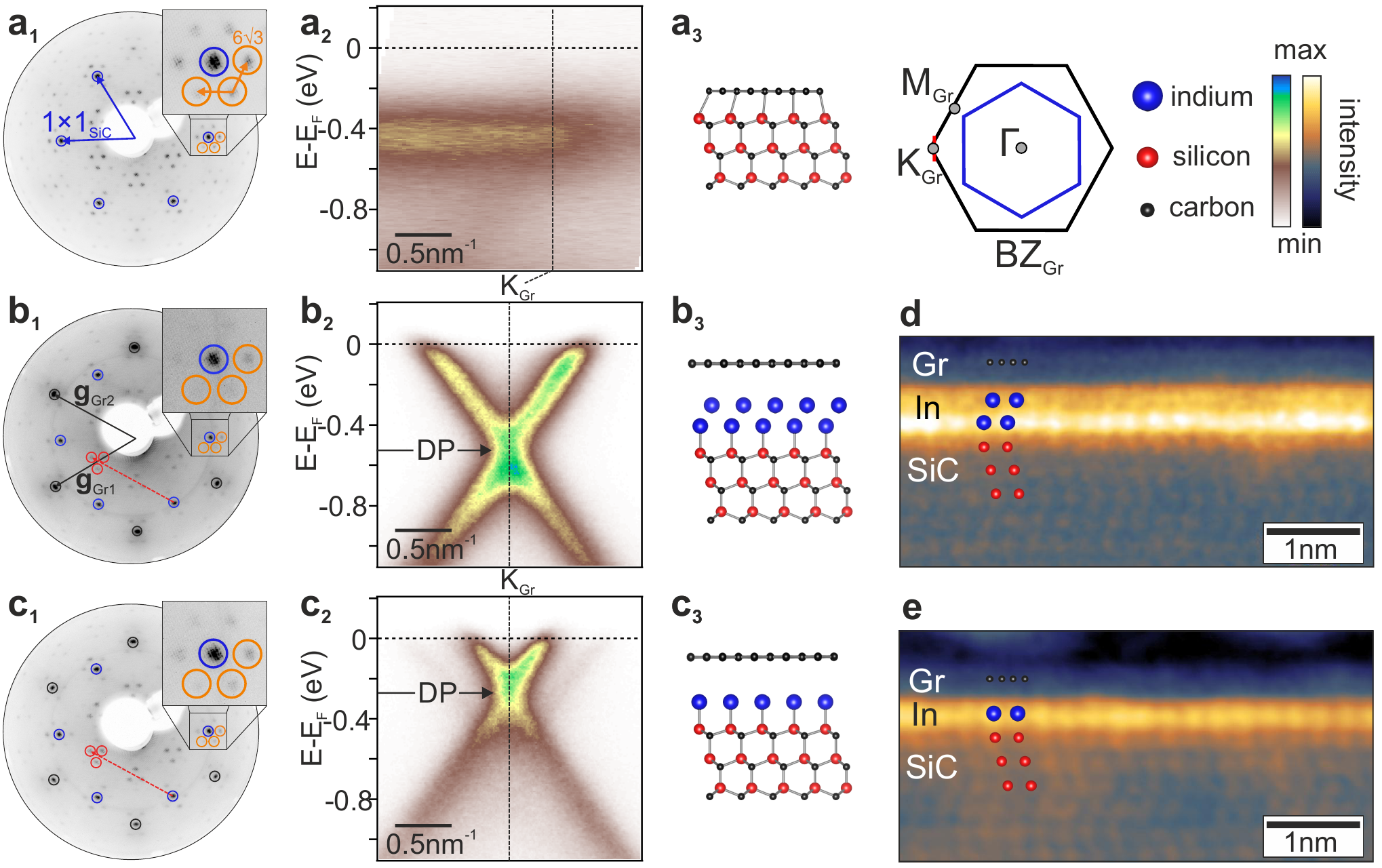}
\caption{\textbf{
The indenene intercalation process.} (a\textsubscript{1},b\textsubscript{1},c\textsubscript{1}) LEED images taken at 100$\,$eV, 
(a\textsubscript{2},b\textsubscript{2},c\textsubscript{2}) ARPES spectra around
the graphene $\text{K}$-point K$_{\text{Gr}}$ and 
(a\textsubscript{3},b\textsubscript{3},c\textsubscript{3}) schematic ball-and-stick model
for \textbf{a} ZLG, \textbf{b} after indium bilayer intercalation and \textbf{c}
graphene capped indenene. LEED images are normalized to
SiC(0001) $(1 \, \times \,1)$ spot (blue) intensities and show diffraction spots
of ($6\sqrt{3} \, \times \,6\sqrt{3}$)R30$^\circ$ periodicity (orange)
\cite{Riedl_2010} and graphene (black). Red marks indicate a selection of
possible scattering vectors between SiC(0001) $(1 \, \times \,1)$ and graphene
\cite{Polley_2019}. \textbf{d,e} Room-temperature STEM image of the 2\;ML and
1\;ML indium intercalated systems, respectively.
}   
\label{fig:StructureIntercalation}
\end{figure*}

Intercalating indenene into the graphene/SiC interface,  the massive
Dirac bands are found to be well preserved, yet, significantly depopulated with $E_F$
shifted from $n$- to $p$-type indenene (Fig.~\ref{fig:DiracXSWXPS}b). 
At the same time, the slight increase in the measured bonding distance d$_{\text{In-Si}}$ indicates
that graphene pulls indenene \textit{away} from the substrate and thus points to 
a further reduction of the ISB strength $\lambda_{\text{ISB}}$ with respect to the pristine case, preserving SOC
as the dominating factor, and stabilizing the topological gap along with the
QSHI state (Fig.~\ref{fig:DiracXSWXPS}c, yellow data point).

The virtue of this graphene-covered QSHI relies on its resilience against
oxidation, which we study by controlled oxygen exposure and subsequent X-ray
photolelectron spectroscopy (XPS) on the pristine (Fig.~\ref{fig:DiracXSWXPS}d)
and intercalated indenene (Fig.~\ref{fig:DiracXSWXPS}e). In both cases, the
as-grown material (black spectra) reveals identical In $3d_{3/2}$/$d_{5/2}$
doublets. Exposing pristine indenene to large doses of oxygen (red) causes
these peaks to broaden and display a chemical shift to higher binding energies,
indicating strong indium oxidation \cite{Lin_1977}. 
In contrast, exposing intercalated indenene
to the same dose of pure oxygen (red), ambient air (green) or even water (blue) has
virtually no impact on the indium oxidation state as well as its band structure (shown in Fig.~4c,d after comprehensive discussion) and thus confirms the protective
function of the graphene overlayer.

%% file: ResultsandDiscussion.tex


\begin{figure*}[ht!]
\includegraphics[width=17.94cm, keepaspectratio]{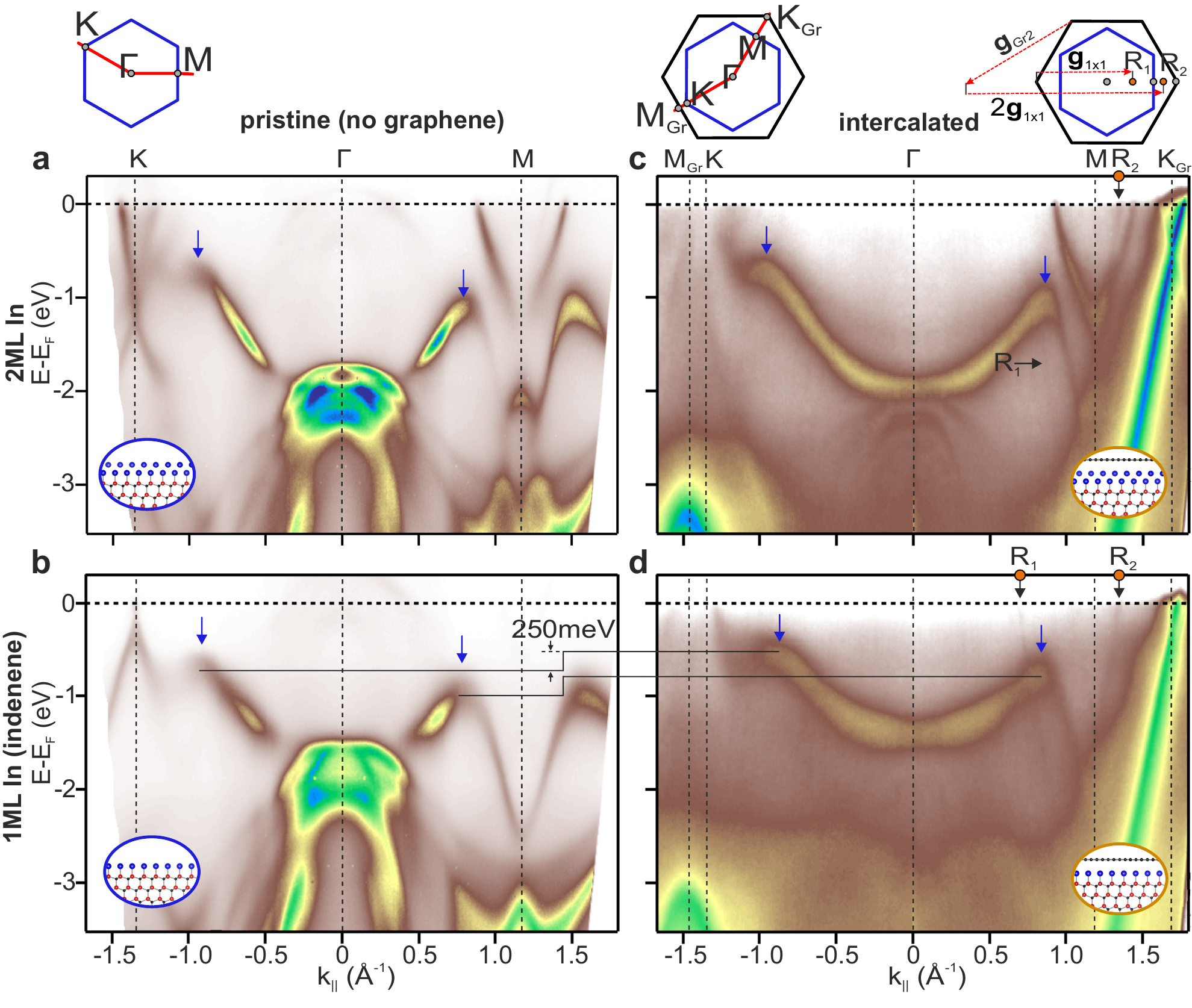}
\caption{
\textbf{The band structure of pristine and graphene capped In phases.} 
ARPES of \textbf{a} pristine bilayer, \textbf{b} monolayer indium and \textbf{c} intercalated bilayer, \textbf{d} monolayer indium on SiC(0001). The pristine data \textbf{(a,b)}
was taken at RT with $h\nu$=21.2$\,$eV, while the intercalated data
\textbf{(c,d)} was taken at $h\nu$=46$\,$eV. Blue arrows indicate distinct band
maxima due to out-of-plane mirror symmetry breaking and orbital hybridization. The top
row illustrations show the Brillouin zones of indenene (blue) and graphene
(black) and the high symmetry $k$-path (red) along which the ARPES data is
shown. Graphene band replicas in \textbf{(c,d)} that are consistent with
electron diffraction off the underlying In/SiC lattice
(sketch top right, i.e., with scattering vector
$\vec{g}_{1\times1}$) are labelled R$_1$, while replicas consistent with
multiple scattering (i.e., with scattering vectors $\vec{g}_{\text{Gr1,2}} +
2\vec{g}_{1\times1}$, $\vec{g}_{\text{Gr1,2}}$ being the graphene reciprocal
lattice vectors) are labelled R$_2$.
}
\label{fig:ARPES}
\end{figure*}

Having summarized the phenomenology, let us now focus on detailed aspects of indium intercalation \cite{Briggs2020,Kim,Hu}, especially the large area growth of monolayer indenene and the identification of its non-trivial topology. Following the well-established recipe
of Ref.\;\onlinecite{Riedl_2010}, the synthesis is initiated by sublimating the
topmost Si atoms off the SiC(0001) substrate, leaving a C-rich buffer layer
referred to as zero-layer graphene (ZLG,
Fig.~\ref{fig:StructureIntercalation}a). Low energy electron diffraction (LEED,
Fig.~\ref{fig:StructureIntercalation}a\textsubscript{1}) reveals a
characteristic ($6\sqrt{3} \, \times \,6\sqrt{3}$)R30$^\circ$ fingerprint
relative to the SiC(0001) ($1 \, \times \, 1$) surface unit
cell signalling a homogeneous ZLG coverage of the substrate. Corresponding
ARPES measurements (Fig.~\ref{fig:StructureIntercalation}a\textsubscript{2})
show a broad and non-dispersive valence state at the $\text{K}$-point of the
graphene Brillouin zone, affirming the covalent bonding
(Fig.~\ref{fig:StructureIntercalation}a\textsubscript{3}) to the underlying SiC
that prevents ZLG from developing linear $\pi$-bands \cite{Riedl_2010}.

In a cyclic process of indium deposition and subsequent annealing, we replace
the top layer carbon for indium as bonding partner to the substrate, hereby
lifting the ZLG template from the subjacent SiC to form quasi-freestanding
monolayer graphene (QFMG). The ($6\sqrt{3} \, \times \,6\sqrt{3}$)R30$^\circ$
LEED signature of ZLG weakens significantly, while the diffraction spots of QFMG
intensify and mark its decoupling from the substrate
(Fig.~\ref{fig:StructureIntercalation}b\textsubscript{1} and inset). ARPES now
exhibits the characteristic $\pi$-band crossing of graphene
(Fig.~\ref{fig:StructureIntercalation}b\textsubscript{2}), with the Dirac point (DP)
at $\text{K}$ (-0.5$\,$eV) lying at slightly ($\sim$80$\,$meV) higher energies as
compared to non-intercalated graphene on a ZLG buffer layer \cite{Riedl_2010}.
Representative scanning transmission electron micrographs (STEM)
(Fig.~\ref{fig:StructureIntercalation}d) show two projected indium layers, each
containing one In atom per Si site of the SiC surface. This puts forward a $(1
\, \times \,1)$ adsorption geometry of 2$\,$ML In on SiC(0001) as illustrated in
Fig.~\ref{fig:StructureIntercalation}b\textsubscript{3} and is corroborated by
the absence of higher order diffraction spots in LEED
(Fig.~\ref{fig:StructureIntercalation}b\textsubscript{1}).

At this point, our experimental results are reminiscent of recent experimental
work on 2$\,$ML In intercalated into a graphene-SiC interface with three
graphene top layers instead of one \cite{Briggs2020}. By thermal removal of In
(550$\, ^{\circ}$C, 30$\,$min) we now convert indium into monolayer indenene,
pushing this material into the QSHI phase. STEM in
Fig.~\ref{fig:StructureIntercalation}e exposes the indenene/graphene sandwich,
with In adsorbed directly above the topmost Si species of SiC
(Fig.~\ref{fig:StructureIntercalation}c\textsubscript{3}) \cite{Bauernfeind}.
The corresponding LEED in
Fig.~\ref{fig:StructureIntercalation}c\textsubscript{1} again shows $(1 \,
\times \,1)$ periodicity, clear-cut evidence for intercalated indenene to
arrange in the same triangular lattice as its pristine counterpart. Quite
remarkably, the transformation from 2$\,$ML indium to 1$\,$ML indenene
redistributes charge within the heterostructure, shifting the graphene Dirac
bands by 0.28\;eV higher in energy compared to the 2$\,$ML film as revealed by
ARPES. (Fig.~\ref{fig:StructureIntercalation}b\textsubscript{2} and
c\textsubscript{2}). Similar band fillings upon post-annealing were already
observed in Ref.\;\onlinecite{Kim}, yet, were tentatively assigned to incomplete
intercalation and a 2$\,$ML phase, respectively. In contrast, we identify
graphene doping as clear-cut fingerprint to distinguish bi- from monolayer
indium.


To further elaborate on this link, we recapitulate the experimental band structures
of pristine 1$\,$ML and 2$\,$ML In films in Fig.~\ref{fig:ARPES}(a,b)
\cite{Erhardt}, and compare them to their intercalated counterparts in
Fig.~\ref{fig:ARPES}(c,d). ARPES at the $\Gamma$-point of both pristine phases
(a,b) shows the intense SiC valence band maximum as well as an indium-related
band dispersing upwards in energy, the latter exposing distinct maxima along the
$\Gamma \text{M}$ and $\Gamma \text{K}$ paths (blue arrows). These maxima arise
from the substrate-induced breaking of mirror symmetry that fosters hybridization
between in- and out-of-plane In $p$ orbitals \cite{Bauernfeind}, and gaps out
the metallic $sp$ bands observed close to the $\text{M}$-point of 2$\,$ML In,
exposing the Dirac states at $\text{K}$ that we discussed in
Fig.~\ref{fig:DiracXSWXPS}a \cite{Erhardt}.

Albeit considerably damped and broadened by the graphene cover, we recognize all
ARPES features of the pristine phases to reappear in their intercalated
counterparts of Fig.~\ref{fig:ARPES}(c,d). Here, we chose a different photon energy at which the indium 2D bands around  $\Gamma$ are more prominent while SiC bands appear suppressed. 
We also identify the intense
$\pi$-band characteristic of graphene, faintly replicated by final state
scattering via higher order reciprocal lattice vectors -- an effect known from
well-ordered intercalated materials \cite{Au_Intercalation} that is also seen in
our LEED data, see red marks in Fig.~\ref{fig:StructureIntercalation} b$_1$,
c$_1$ and Ref. \cite{Polley_2019}.

With respect to pristine indenene, the intercalated indenene bands are shifted
up by approximately 250\;meV (horizontal lines in Fig.~\ref{fig:ARPES}),
corresponding to an overall charge carrier depletion of $\Delta n_{\text{In}}
\approx -3\times10^{12}$ carriers per cm$^{2}$ that we estimate from the 
Fermi surface area (see Methods) and is in qualitative
agreement with the n- to p-type transition in Fig.~\ref{fig:DiracXSWXPS}(a,b).
At the same time, Fig.~\ref{fig:StructureIntercalation}c\textsubscript{2}
suggests the graphene layer to accumulate $\Delta n_{\text{Gr}}\approx+3.4
\times10^{12}$\;cm$^{-2}$ with respect to charge neutral graphene, indicating a
dominating electron transfer from indenene to graphene that is 
balanced by the underlying SiC substrate \cite{Mammadov_2014}.



 \begin{figure}[ht]
\includegraphics[width=8.66cm, keepaspectratio]{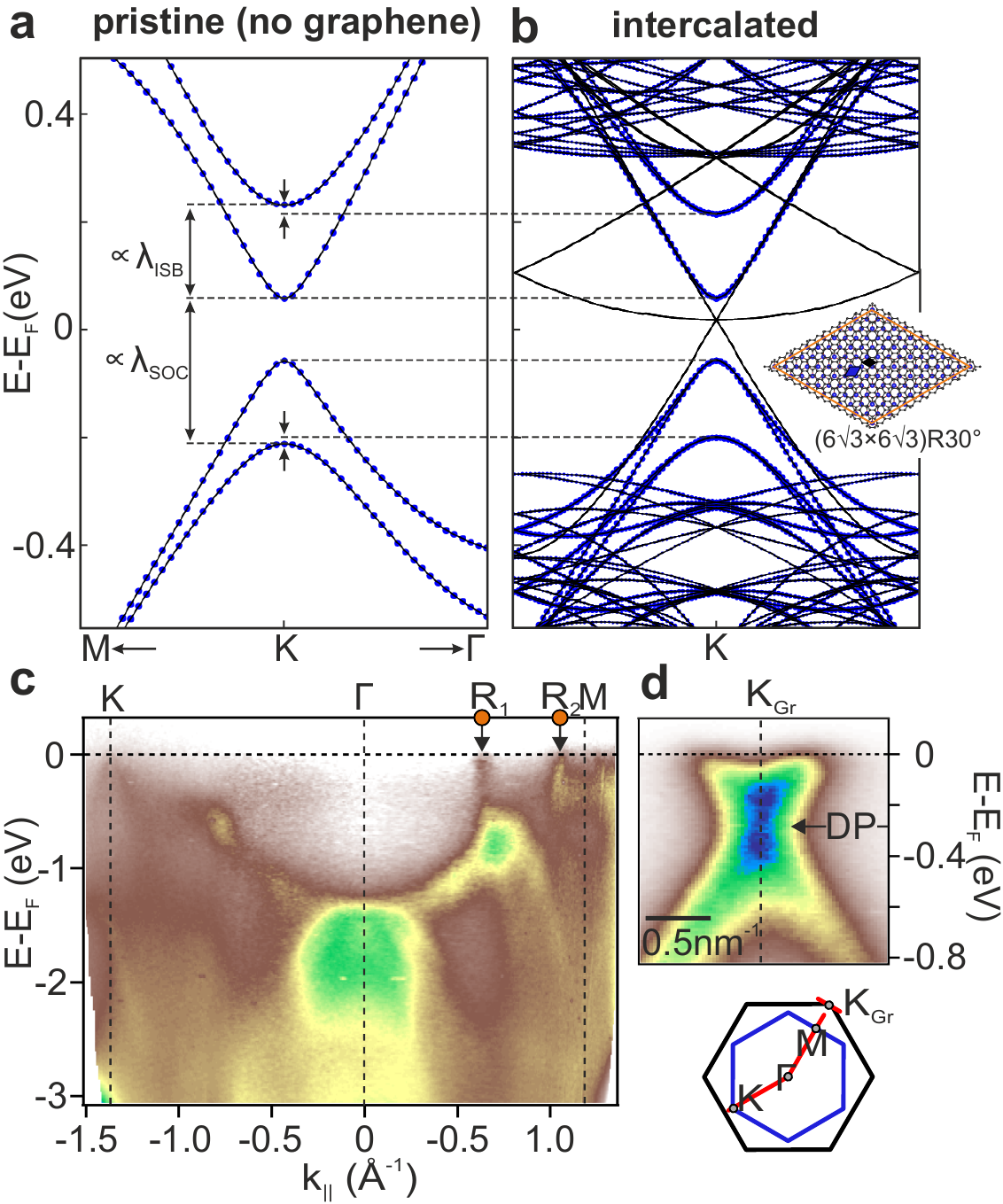}
\caption{
\textbf{Indenene band splitting at K as QSHI signature.} 
DFT (PBE) band structure (black) around the $\text{K}$-point of pristine
\textbf{a} and intercalated indenene \textbf{b} with indium character
represented by the blue disk size (radius). Band splitting $among$ the valence
(conduction) bands is driven by the ISB strength $\lambda_{\text{ISB}}$ (arrow),
while the larger $\lambda_{\text{SOC}}$ splitting (arrow) opens a
topologically non-trivial gap. Note that the combined ($6\sqrt{3} \, \times
\,6\sqrt{3}$)R30$^\circ$ super cell (inset) leads to band backfolding which
projects graphene bands (black) into the indenene gap. 
Importantly, in order to disentangle the role of d$_{\text{In-Si}}$ from the graphene-induced change in $\lambda_{\text{ISB}}$, we calculate both band structures at the In-Si bonding distance determined from XSW of intercalated indenene.
\textbf{c} Normalized ARPES
measurements of the $\text{K}\Gamma \text{M}$-direction of intercalated indenene and \textbf{d} the graphene K-point, both taken at RT and $h\nu$=21.2$\,$eV after immersion in liquid water and mild degassing (see Methods).
}
\label{fig:XSW_DFT}
\end{figure}

Irrespective of the exact mechanism, the observed charge transfer indicates
non-negligible interaction between indenene and graphene that could potentially
affect indenene's topology, 
for instance by modulating $\lambda_{\text{ISB}}$. On top of that, the very presence of graphene tends to increase the distance d$_{\text{In-Si}}$ (see Fig.~\ref{fig:DiracXSWXPS}c) hence enhancing the dominance of SOC over ISB.
To quantify these effects, we add one relevant parameter to our analysis, namely the distance between indium and graphene d$_{\text{In-Gr}}$. 
Both d$_{\text{In-Si}}$ and d$_{\text{In-Gr}}$ can experimentally be accessed by
X-ray standing wave (XSW) photoemission that utilizes Bragg scattering off the
bulk substrate lattice to generate a modulated wave-field along the surface
normal. The photoelectron yields of the In and C core levels get strongly
enhanced when the field maxima coincide with the corresponding atomic layer (for
details see Methods) \cite{Woodruff_2005}. Tuning the photon energy and thus the
phase of the standing wave, we measured the layer distances of intercalated
indenene and found d$_{\text{In-Si}}=(2.74 \pm 0.04)\, \text{\AA}$ and
d$_{\text{In-Gr}}=(3.35\pm0.04) \, \text{\AA}$. 
The latter amounts to about 92$\%$ of
indium's van der Waals bonding length \cite{Mantina_2009}, underlining
graphene's quasi-freestanding character \cite{Lin_2022}. 
In order to disentangle the effect on $\lambda_{\text{ISB}}$ induced by the SiC-substrate underneath from that of the graphene capping on top, we perform density functional theory (DFT) calculations for indenene without and with graphene
cover as shown in Fig.~\ref{fig:XSW_DFT}a and b for the same d$_{\text{In-Si}}$.
The results clearly indicate that the bands at the valley momenta preserve their non-trivial sequence. The splitting due to SOC strength $\lambda_{\text{SOC}}$ gets slightly renormalized but is roughly compensated by a concomitant reduction of the $\lambda_{\text{ISB}}$, effectively leaving the band gap unchanged (see arrows).
For this reason we conclude that the variation of the In-Si distance d$_{\text{In-Si}}$ induced by the presence of graphene is the main factor determining $\lambda_{\text{ISB}}$, as indicated in Fig.~\ref{fig:DiracXSWXPS}c.
This completes the proof of intercalated indenene to be topologically
robust and clearly placed within the QSHI regime. 

Combined with its remarkable resilience against ambient conditions, this opens up a wealth of
experimental possibilities to characterize and manipulate this 2D topological
insulator \textit{ex-situ}. While the conductive nature of the graphene cap may
still interfere with meaningful edge transport measurements, its effective
protection of the monolayer QSHI certainly paves the way for (nano)fabrication
of device structures, e.g., for field effect gating. It also greatly facilitates
optical or infrared experiments such as Raman and Landau level spectroscopy in
non-vacuum settings, techniques that will give further insight to the physics of
this intriguing QSHI. We underline that -- quite remarkably -- intercalated
indenene remains inert even upon immersion in water (see
Figs.~\ref{fig:DiracXSWXPS}e and \ref{fig:XSW_DFT}). Beyond the specific case of
indenene, this suggests graphene intercalation as a promising route towards the
application of atomic monolayers as functional quantum materials, even in rough
chemical environments \cite{Glavin2020}.